# The Massive Neutrino-like Particle of the Non-linear Electromagnetic Field Theory

Alexander G. Kyriakos

*Saint-Petersburg State Institute of Technology,*
*St.Petersburg, Russia*
*Present address: Athens, Greece, e-mail: agkyriak@yahoo.com*

## Abstract

A special nonlinear electromagnetic field theory - the curvilinear wave electrodynamics (CWED), whose equation in the matrix form is similar to the equation of the Dirac lepton theory, is offered. It is shown, that the solution of this equation is the motion of the circularly polarized electromagnetic wave on a circular trajectory. It is also shown, that such wave can be considered as a massive neutral particle with half spin, similar to neutrino. This theory is mathematically concurrent to the neutrino theory of the Standard Model, but does not have the contradictions of Standard Model with last data of neutrino experiments.

PASC 12.10.-g  Unified field theories and models.
PASC 12.90.+b  Miscellaneous theoretical ideas and model.

## 1.0.  Introduction. Neutrino of Standard Model theory

Before we state the non-linear electromagnetic theory of massive neutrino-like particle, we will briefly look into the present theory of neutrino.

The present status of the problem of neutrino theory is summarized in [1,2]. The theory of electroweak interactions including neutrinos combined with the Quantum Chromo-Dynamics (QCD) is now called the Standard Model (SM).

## 1.1. Neutrinos features in SM theory

In the Standard Model neutrinos are strictly massless, $m = 0$; all neutrinos are left-handed, helicity -1, and all antineutrinos are right-handed, helicity +1; lepton family number is strictly conserved. But modern experimental evidence indicates that all of these statements are in fact doubtful [1].

In the Standard Model theory, neutrino and antineutrino have opposite helicity. It is mathematically possible that this is in fact the *only* difference between neutrinos and antineutrinos, i.e. a right-handed "neutrino" would be an antineutrino. Particles of this sort are called Majorana particles.

As long as the neutrino is massless, its helicity is completely defined, and a Majorana neutrino would be a different particle from its antineutrino. But if neutrinos have mass, and therefore do not travel at exactly the speed of light, it is possible to define a reference frame in which the helicity would be flipped. This means that there is effectively a mixing between the neutrino and the antineutrino (violating lepton number conservation).

### 1.1.1. Helicity and Chirality

We will use the original quantum form of the Dirac lepton equation. There are two bispinor conjugate Dirac equation forms [3]:

$$\left[ \left( \hat{\alpha}_o \hat{\varepsilon} + c \hat{\vec{\alpha}} \cdot \hat{\vec{p}} \right) + \hat{\beta} \, mc^2 \right] \psi = 0 \,, \tag{1.1}$$

$$\psi^+ \left[ \left( \hat{\alpha}_o \hat{\varepsilon} - c \hat{\vec{\alpha}} \cdot \hat{\vec{p}} \right) - \hat{\beta} \, mc^2 \right] = 0 \,, \tag{1.2}$$

where $\hat{\varepsilon} = i\hbar \dfrac{\partial}{\partial t}, \hat{\vec{p}} = -i\hbar \vec{\nabla}$ are the operators of the energy and momentum, $\varepsilon, \vec{p}$ are the electron energy and momentum, $c$ is the light velocity, $m$ is the electron mass, and $\hat{\alpha}_o = \hat{1}, \hat{\vec{\alpha}}, \hat{\alpha}_4 \equiv \hat{\beta}$ are the original set of Dirac's matrices:

$$\hat{\alpha}_0 = \begin{pmatrix} \hat{\sigma}_0 & 0 \\ 0 & \hat{\sigma}_0 \end{pmatrix}, \; \hat{\vec{\alpha}} = \begin{pmatrix} 0 & \hat{\vec{\sigma}} \\ \hat{\vec{\sigma}} & 0 \end{pmatrix}, \; \hat{\beta} \equiv \hat{\alpha}_4 = \begin{pmatrix} \hat{\sigma}_0 & 0 \\ 0 & -\hat{\sigma}_0 \end{pmatrix}, \; \hat{\alpha}_s = i \begin{pmatrix} 0 & -\hat{\sigma}_0 \\ \hat{\sigma}_0 & 0 \end{pmatrix}, \tag{1.3}$$

where $\hat{\vec{\sigma}}$ are Pauli spin matrices: $\hat{\sigma}_1 = \begin{pmatrix} 0 & 1 \\ 1 & 0 \end{pmatrix}, \; \hat{\sigma}_2 = \begin{pmatrix} 0 & -i \\ i & 0 \end{pmatrix}, \; \hat{\sigma}_3 = \begin{pmatrix} 1 & 0 \\ 0 & -1 \end{pmatrix}, \; \hat{\sigma}_0 = \begin{pmatrix} 1 & 0 \\ 0 & 1 \end{pmatrix};$

$\psi$ is the wave function $\psi = \begin{pmatrix} \psi_1 \\ \psi_2 \\ \psi_3 \\ \psi_4 \end{pmatrix}$ called bispinor, $\psi^+$ is the Hermitian-conjugate wave function.

*Helicity* refers to the relation between a particle's spin and direction of motion. To a particle in motion is associated the axis defined by its momentum, and its helicity is defined by the projection of the particle's spin $\vec{s}$ on this axis: the helicity is $h = \dfrac{\vec{s} \vec{p}}{|\vec{s}||\vec{p}|}$, i.e. the component of angular momentum along the momentum.

The helicity operator thus projects out two physical states, with the spin along or opposite the direction of motion - whether the particle is massive or not. If the spin is projected parallel on the direction of motion, the particle is of right helicity, if the projection is antiparallel to the direction of motion, the particle has left helicity.

Something is *chiral* when it cannot be superimposed on its mirror image, like for example our hands. Like our hands, chiral objects are classified into left-chiral and right-chiral objects. For a massless fermion, the Dirac equation reads

$$\alpha^\mu \partial_\mu \psi = 0, \ \ \mu = 1,2,3,4 \,, \tag{1.4}$$

which is also satisfied by $\alpha_5 \psi$ :

$$\alpha^\mu \partial_\mu (\alpha_5 \psi) = 0 \,, \tag{1.5}$$

where the combination of the $\alpha$ matrices, $\alpha_5 = \alpha_0 \alpha_1 \alpha_2 \alpha_3$ has the properties $\alpha_5^2 = 1$ and $\{\alpha_5, \alpha_\mu\} = 0$ . This allows us to define the *chirality operators* which project out left-handed and right-handed states:

$$\psi_L = \frac{1}{2} \left( 1 - \alpha_5 \right) \psi \ \ \text{and} \ \ \psi_R = \frac{1}{2} \left( 1 + \alpha_5 \right) \psi \,, \tag{1.6}$$

where $\psi_L$ and $\psi_R$ satisfy the equations $\alpha_5 \psi_L = -\psi_L$ and $\alpha_5 \psi_R = \psi_R$ , so the chiral fields are eigenfields of $\alpha_5$ , regardless of their mass.

We can express any fermion as $\psi = \psi_L + \psi_R$ , so that a massive particle always has a *L*-handed as well as a *R*-handed component. In the massless case $\psi$ however "disintegrates" into separate helicity states: the Dirac equation splits into two independent parts, reformulated as the <u>Weyl equations</u>

$$\frac{\hat{\vec{\sigma}} \hat{\vec{p}}}{\left| \hat{\vec{\sigma}} \right| \left\| \hat{\vec{p}} \right\|} \left[ \frac{1}{2} \left( 1 \pm \alpha_5 \right) \psi \right] = \pm \frac{1}{2} \left( 1 \pm \alpha_5 \right) \psi \,, \tag{1.7}$$

where $\dfrac{\hat{\vec{\sigma}} \hat{\vec{p}}}{\left| \hat{\vec{\sigma}} \right| \left\| \hat{\vec{p}} \right\|}$ is the helicity operator expressed in terms of the Pauli spin matrices $\hat{\vec{\sigma}}$ .

The Weyl fermions, i.e. the massless chiral states $\frac{1}{2} \left( 1 \pm \alpha_5 \right) \psi$ , are physical since they correspond to eigenstates of the helicity operator. A massless particle, which is in perpetual motion, thus has an unchangeable helicity. The reason is that its momentum cannot be altered, and its spin of course remains unchanged.

For a massive particle however, we can perform a Lorentz transformation along the direction of the particle's momentum with a velocity larger than the particle's, changing the direction of the momentum. Since the spin direction remains the same, the helicity of the particle changes.

### 1.1.2. Electromagnetic characteristics neutrino

It is interesting, that, in spite of neutrality, neutrino possesses electromagnetic characteristics. The analysis of these characteristics allows us to conclude about the nature of neutrino mass [4]. Electromagnetic properties of Dirac's and Majorano's neutrino appear to be essentially various. Dirac's massive neutrino as a result of the account of interaction with vacuum receives the magnetic moment. And, the neutrino magnetic moment is directed lengthways a spin, and the magnetic moment antineutrino - against a spin. Thus, the particle and the antiparticle differ by the direction of the magnetic moment. For massive Majorana neutrino, identical to its antiparticle, it appears, that it cannot have neither magnetic nor electric moment.

It also appeared, that the mass and the magnetic moment of neutrino are complex nonlinear functions of field strength and energy of a particle.

Moving in an external field, alongside with the magnetic moment, the Dirac neutrino gets as well the dipole electric moment $d_\nu$. Calculations show, that the electric moment of massive Dirac's neutrino, moving in a constant external general view field, is proportional to a pseudo-scalar $(\vec{E} \cdot \vec{H})$, which changes sign at the reflection of time. Thus, the electric moment is induced by an external field, if for this field, the pseudo-scalar $(\vec{E} \cdot \vec{H}) \neq 0$ and its existence does not contradict to T-invariancy of Standard Model. In other words the dipole electric moment of Dirac's neutrino, as well as the magnetic, has dynamic nature.

Note also that there is one electromagnetic characteristic of Dirac's neutrino, which takes place also for Majorano's neutrino: the anapole (or toroidal dipole) moment.

Below we will show that in the framework of the non-linear electromagnetic theory the massive neutrino has the conserved inner (poloidal) helicity, owing to which the above features occur, and is fully described by the Dirac-like lepton equation.

## 2.0. Neutrino-like particles in the electrodynamics of the curvilinear waves

In the previous papers [5,6] on base of the some non-linear electromagnetic theory we have considered the electromagnetic representation of the quantum theory of electron and obtained the Dirac-like equation for electron-like particles. In the framework of this theory, which can also be named the curvilinear wave electrodynamics (briefly CWED), the electron-like particles are the twirling *plane-polarized* semi-photon-like particles.

In the present paper we will show that the solution of this Dirac-like equation describes also the motion of the *circular-polarized* semi-photon-like particle on a circular trajectory. It is shown that such particle can be considered as a neutral particle with half spin, similar to neutrino.

## 3.0. Plane and circularly polarized electromagnetic waves

Electromagnetic waves emitted by charged particles are in general circularly (or elliptically) polarized. Electromagnetic waves are also transverse in the sense that associated electric and magnetic field vectors are both perpendicular to the direction of wave propagation.

Circularly polarized waves carry energy $\varepsilon$ and momentum $\vec{p}$ as well as angular momentum $\vec{J}$, which are defined by energy density $U = \frac{1}{8\pi}\left(\vec{E}^2 + \vec{H}^2\right)$, momentum density $\vec{g} = \frac{1}{c^2}\vec{S}_P$ and angular momentum flux density, which is given by

$$\vec{s} = \vec{r} \times \vec{g} = \frac{1}{4\pi}\frac{1}{c} r \times \vec{E} \times \vec{H} , \qquad (3.1)$$

where $\vec{S}_P = \frac{c}{4\pi}\vec{E} \times \vec{H}$ is the Poynting vector indicates not only the magnitude of the energy flux density, but also the direction of energy flow. For simple electromagnetic waves, the Poynting vector is in the same direction as the wavevector.

The angular momentum flux density can be checked by the circularly motion of the electron in the circularly polarized wave field [7].

A plane electromagnetic wave (as well as the waves with other polarizations) can be considered as vector combination of two or more circularly polarized waves rotating in opposite directions. The figure 1 below shows propagation of electric field associated with a circularly polarized wave with positive (right) and negative (left) helicity.

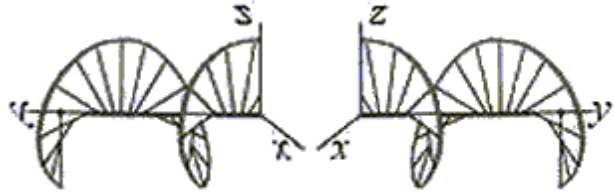

Fig. 1

Positive helicity is the case such that a right screw would move in the direction of wave propagation if rotated with the electric field (note that in optics, it is called "left hand" circular polarization). Negative helicity (right hand polarization in optics)) refers to rotation in the opposite direction. The direction of the end of the helix indicates the head of the electric field vector, which is rotating about the *y* axis.

Since it is impossible by any transformation, except for the spatial reflection, to transfer the right (left) spiral to the left (right) spiral, the *circular polarization of photons is their integral characteristic kept at all transformations, except of the mirror transformation*.

Since the photon helicity is connected to the field rotation, in classical electrodynamics they also talk about rotation of a photon and they enter the photon rotation characteristic – the angular momentum or spin of a photon. In quantum mechanics the attributing to a photon of the spin has some conditional character, as the spin is named the internal angular momentum of a particle in those systems, in which the considered particle is in rest. Therefore in case of a photon, whose speed can not be other than the light speed, it is correct to talk more about the photon helicity than about the spin [8]. In this case it is possible to define as helicity the vector [7]

$$\vec{h}_{ph} \equiv \vec{s}_{ph} = \pm \frac{\varepsilon_{ph}}{\omega} \vec{p}^{\,0} \quad , \tag{3.2}$$

where $\vec{p}^{\,0}$ is the unit Pointing vector, $\varepsilon_{ph}$ and $\omega$ are the photon energy and circular frequency correspondingly. Apparently the angular momentum value of this vector is equal to $\left| \vec{h}_{ph} \right| = 1\hbar$.

Then according to our hypothesis the p-helicity vector of neutrino, as twirled semi-photon, should have tangential direction to the curvilinear trajectory of twirled wave motion, and its angular momentum should be equal to half of the above value $\left| \vec{h}_{\nu} \right| = \frac{1}{2}\hbar$.

## 4.0.  Quantum form of the circularly polarized electromagnetic wave equations

Obviously, by the twirling of the circular polarized photon in a ring, its helicity does not disappear, but become poloidal (in torus)  helicity (p-helicity). At the same time, the movement of fields of a photon along a circular trajectory forms other characteristics of an elementary particle - namely the angular momentum of a particle, or spin. Apparently, the spin of a massive particle and its poloidal angular momentum (p-helicity) are different characteristics. Since these characteristics are the own internal characteristics of a photon, *in the non-linear electromagnetic theory the spin and the poloidal helicity of a particle are independent and conserved values.*

We will show that there is the Dirac-like equation, which can be considered as the equation of the twirled circularly polarized waves.

Let us consider the plane electromagnetic wave moving, for example, along $y$ - axis:

$$\begin{cases} \vec{E} = \vec{E}_o e^{-i(\omega t \pm ky)}, \\ \vec{H} = \vec{H}_o e^{-i(\omega t \pm ky)}, \end{cases} \tag{4.1}$$

In general case the electromagnetic wave, moving along $y$ - axis, has two polarizations and contains the following four field vectors:

$$(E_x, E_z, H_x, H_z) \tag{4.2}$$

As in this case $E_y = H_y = 0$ for all transformations, we can compare the set (4.2) with the Dirac wave function. Let's now consider the electromagnetic wave equation:

$$\left( \frac{\partial^2}{\partial t^2} - c^2 \vec{\nabla}^2 \right) \vec{F} = 0, \tag{4.3}$$

where $\vec{F}$ is whichever of the electromagnetic wave functions, particularly, the fields (4.2). In other words this equation represents four equations for each wave function of the electromagnetic field.

In case of the wave, moving along the $y$ - axis, we can write this equation in the following form:

$$\left( \hat{\varepsilon}^2 - c^2 \hat{\vec{p}}^2 \right) \vec{F}(y) = 0, \tag{4.4}$$

The equation (4.4) can also be represented in the form of the Klein-Gordon equation [3] without mass:

$$\left[ \left( \hat{\alpha}_o \hat{\varepsilon} \right)^2 - c^2 \left( \hat{\vec{\alpha}} \ \hat{\vec{p}} \right)^2 \right] \Psi = 0, \tag{4.5}$$

where $\Psi$ is a matrix, which in some way consists of the fields (4.2). In fact, taking into account that $\left( \hat{\alpha}_o \hat{\varepsilon} \right)^2 = \hat{\varepsilon}^2$, $\left( \hat{\vec{\alpha}} \ \hat{\vec{p}} \right)^2 = \hat{\vec{p}}^2$, we can realize that the equations (4.4) and (4.5) are equivalent.

Taking into account that in case of photon $\omega = \varepsilon / \hbar$ and $k = p / \hbar$ and using (4.1), from (4.5) we obtain $\varepsilon = cp$, as for a photon. Therefore we can name the wave $\Psi$ the photon-like particle.

Factorizing (4.5) and multiplying it from the left on the Hermitian-conjugate function $\psi^+$ we get:

$$\Psi^+ \left( \hat{\alpha}_o \hat{\varepsilon} - c \hat{\vec{\alpha}} \ \hat{\vec{p}} \right) \left( \hat{\alpha}_o \hat{\varepsilon} + c \hat{\vec{\alpha}} \ \hat{\vec{p}} \right) \Psi = 0, \tag{4.6}$$

The equation (4.6) may be disintegrated on two Dirac equations without mass:

$$\Psi^+ \left( \hat{\alpha}_o \hat{\varepsilon} - c \hat{\vec{\alpha}} \ \hat{\vec{p}} \right) = 0, \tag{4.7}$$

$$\left( \hat{\alpha}_o \hat{\varepsilon} + c \hat{\vec{\alpha}} \ \hat{\vec{p}} \right) \Psi = 0, \tag{4.8}$$

We will name these particles semi-photon particles.

Let us show that if we choose the wave function as (and only as)

$$\Psi = \begin{pmatrix} E_x \\ E_z \\ iH_x \\ iH_z \end{pmatrix}, \quad \Psi^+ = \begin{pmatrix} E_x & E_z & -iH_x & -iH_z \end{pmatrix}, \tag{4.9}$$

the *equations (4.7) and (4.8) are the correct Maxwell equations of the circular polarized electromagnetic wave.* (For all other directions of the electromagnetic waves the matrices choice can be obtained by permutations of indexes [5]).

Putting (4.4) in (4.7) and (4.8) we emerge the Maxwell equation in the case of electromagnetic waves:

$$\begin{cases} \dfrac{1}{c}\dfrac{\partial E_x}{\partial t} - \dfrac{\partial H_z}{\partial y} = 0 \\[2mm] \dfrac{1}{c}\dfrac{\partial H_z}{\partial t} - \dfrac{\partial E_x}{\partial y} = 0 \\[2mm] \dfrac{1}{c}\dfrac{\partial E_z}{\partial t} + \dfrac{\partial H_x}{\partial y} = 0 \\[2mm] \dfrac{1}{c}\dfrac{\partial H_x}{\partial t} + \dfrac{\partial E_z}{\partial y} = 0 \end{cases}, \quad (4.10') \qquad\qquad \begin{cases} \dfrac{1}{c}\dfrac{\partial E_x}{\partial t} + \dfrac{\partial H_z}{\partial y} = 0 \\[2mm] \dfrac{1}{c}\dfrac{\partial H_z}{\partial t} + \dfrac{\partial E_x}{\partial y} = 0 \\[2mm] \dfrac{1}{c}\dfrac{\partial E_z}{\partial t} - \dfrac{\partial H_x}{\partial y} = 0 \\[2mm] \dfrac{1}{c}\dfrac{\partial H_x}{\partial t} - \dfrac{\partial E_z}{\partial y} = 0 \end{cases}, \quad (4.10'')$$

or in the vector form [9,10]:

$$\begin{cases} \dfrac{1}{c}\dfrac{\partial \vec{H}}{\partial t} + rot\ \vec{E} = 0 \\[2mm] \dfrac{1}{c}\dfrac{\partial \vec{E}}{\partial t} - rot\ \vec{H} = 0 \end{cases}, \tag{4.11}$$

The electromagnetic wave equation has the harmonic solution view:

$$\psi_\mu = A_\mu e^{-\frac{i}{\hbar}(\varepsilon t - \vec{p}\vec{r} + \delta)}, \tag{4.12}$$

where $\mu = 1,2,3,4$, $A_j$ are the amplitudes and $\delta$ is the constant phase.

Puting here $A_\mu = A_0$, $\delta = 0$, we obtain the following the trigonometric form of the equation solutions:

$$\begin{cases} E_x = A_0 \cos(\omega t - ky) \\ H_z = -A_0 \cos(\omega t - ky) \\ E_z = -A_0 \sin(\omega t - ky) \\ H_x = -A_0 \sin(\omega t - ky) \end{cases}, \quad (4.13') \qquad\qquad \begin{cases} E_x = A_0 \cos(\omega t - ky) \\ H_z = A_0 \cos(\omega t - ky) \\ E_z = -A_0 \sin(\omega t - ky) \\ H_x = A_0 \sin(\omega t - ky) \end{cases}, \quad (4.13'')$$

Let us show that the vectors $\vec{E}$ and $\vec{H}$ rotate in the *XOZ* plain. Actually, putting $y = 0$ we obtain:

$$\vec{E} = E_x \vec{i} + E_z \vec{k} = A_0\left(\vec{i} \cos\omega t - \vec{k} \sin\omega t\right), \tag{4.14'}$$

$$\vec{H} = H_x \vec{i} + H_z \vec{k} = A_0\left(-\vec{i} \sin\omega t - \vec{k} \cos\omega t\right), \tag{4.14''}$$

and

$$\vec{E} = E_x \vec{i} + E_z \vec{k} = A_0 \left( \vec{i} \cos \omega\, t - \vec{k} \sin \omega\, t \right), \qquad (4.15')$$

$$\vec{H} = H_x \vec{i} + H_z \vec{k} = A_0 \left( \vec{i} \sin \omega\, t + \vec{k} \cos \omega\, t \right), \qquad (4.15'')$$

where $\vec{i}$, $\vec{k}$ are the unit vectors of the *OX* and *OZ* axes. It is not difficult to show by known algebraic analysis [10] that we have obtained the cyclic polarised wave. But to keep in evidence we will analyse these relations from geometrical point of view.

The Poynting vector defines the direction of the wave motion:

$$\vec{S}_P = \frac{c}{4\pi} \vec{E} \times \vec{H} = -\vec{j}\, \frac{c}{4\pi} \left( E_x H_z - E_z H_x \right), \qquad (4.16)$$

where $\vec{j}$ is the unit vectors of the *OY* axis. Calculating the above we have for (4.13') and (4.13''):

$$\vec{S}_P = \frac{c}{4\pi} A_0^2 \vec{j} \quad, \qquad (4.17)$$

and

$$\vec{S}_P = -\frac{c}{4\pi} A_0^2 \vec{j} \quad, \qquad (4.18)$$

correspondingly. Thus, the photons of the right and left systems (4.13') and (4.13'') move in the contrary directions.

Fixing the vector $\vec{E}, \vec{H}$ positions in two successive time instants (in the initial instant ($t = 0$) and through the little time period $\Delta t$), we can define the rotation direction. The results are imaged on the figures 2 and 3 correspondingly:

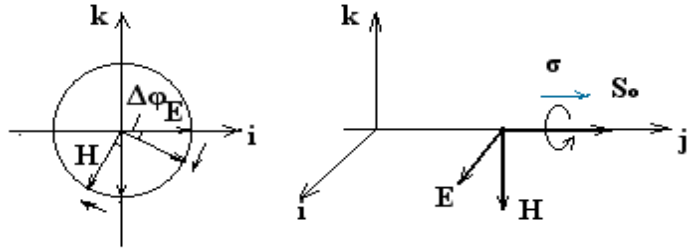

Fig. 2

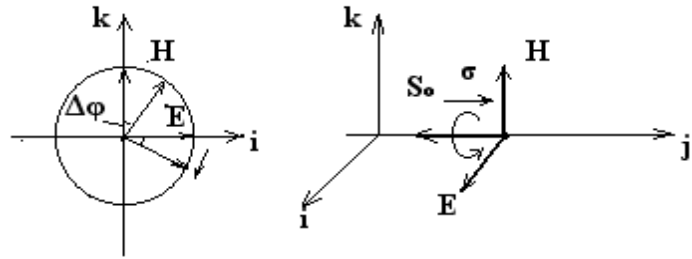

Fig. 3

As we can see the equation sets (4.10') and (4.10'') describe the waves with right and left circular polarization correspondingly.

But the neutrino-like particle, according to the present experimental data, must have a mass. Therefore, it must be described not by the equations (4.7) or (4.8), but by the Dirac-like equation with a mass term.

The question arises: which mathematical transformation can turn the equations (4.7) and (4.8) without mass term into the Dirac equations (1.1) and (1.2), which have the mass term? We have shown [6] that it can be made, at least, in two ways: either by using curvilinear metrics, or by using differential geometry. Here we use only the second way.

## 5.0. Appearance of the mass term

*Let us show that mass term appears when the initial electromagnetic wave changes its trajectory from the linear to the curvilinear*. Noting that the Pauli matrices are the generators of the rotation transformations in the 2D-space, we can suppose that this curvilinear trajectory is plane.

Let the circular-polarized wave (4.1), which have the field vectors $(E_x, E_z, H_x, H_z)$, be twirled with some radius $r_{\kappa}$ in the plane $(X', O', Y')$ of a fixed co-ordinate system $(X', Y', Z', O')$ so that $E_x, H_x$ are parallel to the plane $(X', O', Y')$ and $E_z, H_z$ are perpendicular to it (see fig. 4)

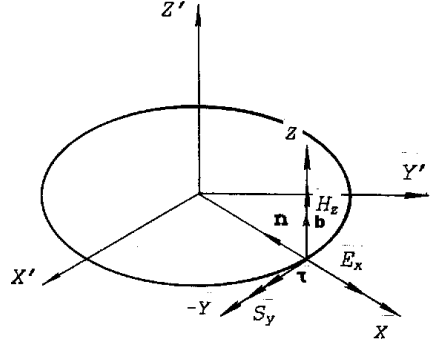

Fig. 4

Consider the expressions

$$\vec{j}^e = \frac{1}{4\pi} \frac{\partial \vec{E}}{\partial t},$$ (5.1')

$$\vec{j}^m = \frac{1}{4\pi} \frac{\partial \vec{H}}{\partial t},$$ (5.1'')

(according to Maxwell [9,10] the expression (5.1') is the displacement electric current; correspondingly we will name the expression (5.1'') the displacement magnetic current).

The above electrical field vector $\vec{E}$, which moves along the curvilinear trajectory can be written in the form:

$$\vec{E}_x = E_x \cdot \vec{n},$$ (5.2)

where $\vec{n}$ is the normal unit-vector of the curve (having direction to the center) and $E_x = \vec{E} \cdot \vec{n}$. The derivative of $\vec{E}_x$ can be represented as:

$$\frac{\partial \vec{E}_x}{\partial t} = \frac{\partial E_x}{\partial t} \vec{n} + E_x \frac{\partial \vec{n}}{\partial t},$$ (5.3)

Here the first term has the same direction as $\vec{E}_x$. The existence of the second term shows that at the twirling of the wave the additional displacement current appears. It is not difficult to show that it has a direction, tangential to the ring:

$$\frac{\partial \vec{n}}{\partial t} = -\upsilon_p \mathrm{K} \vec{\tau} \,, \qquad (5.4)$$

where $\vec{\tau}$ is the tangential unit-vector, $\upsilon_p \equiv c$ is the electromagnetic wave velocity, $\mathrm{K} = \dfrac{1}{r_\mathrm{K}}$ is the curvature of the trajectory and $r_\mathrm{K}$ is the curvature radius. Thus, the displacement current of the plane wave, moving along the ring, can be written in the form:

$$\vec{j}^{\,e} = \frac{1}{4\pi}\frac{\partial E_x}{\partial t}\vec{n} - \frac{1}{4\pi}\omega_\mathrm{K} E_x \cdot \vec{\tau} \,, \qquad (5.5)$$

where $\omega_\mathrm{K} = \dfrac{\upsilon_p}{r_\mathrm{K}} \equiv c\mathrm{K}$ we name the curvature angular velocity, $\vec{j}_n^{\,e} = \dfrac{1}{4\pi}\dfrac{\partial E_x}{\partial t}\vec{n}$ and $\vec{j}_\tau^{\,e} = -\dfrac{\omega_\mathrm{K}}{4\pi}E_x \cdot \vec{\tau}$ are the normal and tangent components of the electric current of the twirled electromagnetic wave, correspondingly.

Thus:

$$\vec{j}^{\,e} = \vec{j}_n^{\,e} + \vec{j}_\tau^{\,e} \,, \qquad (5.6)$$

The currents $\vec{j}_n$ and $\vec{j}_\tau$ are always mutually perpendicular, so that we can write them in the complex form:

$$j^e = j_n^e + ij_\tau^e \,, \qquad (5.7)$$

Following the same calculation it is not difficult to see that the magnetic current appears also:

$$j^m = j_n^m + ij_\tau^m \,, \qquad (5.8)$$

where $\vec{j}_n^{\,m} = \dfrac{1}{4\pi}\dfrac{\partial H_x}{\partial t}\vec{n}$ and $\vec{j}_\tau^{\,m} = -\dfrac{\omega_\mathrm{K}}{4\pi}H_x \cdot \vec{\tau}$ are the normal and tangent components of the magnetic current.

Since for the circular polarized wave

$$\begin{cases} E_x = E_{xo}e^{-i(\omega t \pm ky)} \\ H_x = H_{xo}e^{-i(\omega t \pm ky)} \end{cases}, \qquad (5.9)$$

the tangential currents are alternate.

## 5.1. The Dirac-like equation of the curvilinear electromagnetic theory

Taking into account the previous section results from (4.10) we obtain the twirled semi-photon equations:

$$\frac{\partial \psi}{\partial t} - c\hat{\vec{\alpha}}\,\vec{\nabla}\psi\ - i\hat{\beta}\frac{c}{r_C}\psi = 0 \,, \qquad (5.10')$$

$$\frac{\partial \psi}{\partial t} + c\hat{\bar{\alpha}}\ \vec{\nabla}\psi\ + i\hat{\beta}\frac{c}{r_C}\psi = 0, \qquad (5.10'')$$

where $\psi$- function is not $\Psi$ - function, but they are connected by some transformation, as it will be shown bellow.

Since [3] they are mathematically the same as the Dirac equations (1.1) and (1.2), we will name these equations the Dirac electron-like equation.

Using (4.10), from (1.1) and (1.2) we obtain electromagnetic form of these equations:

$$\begin{cases} \dfrac{1}{c}\dfrac{\partial E'_x}{\partial t} - \dfrac{\partial H'_z}{\partial y} = -ij_x^e \\[2mm] \dfrac{1}{c}\dfrac{\partial H'_z}{\partial t} - \dfrac{\partial E'_x}{\partial y} = ij_z^m \\[2mm] \dfrac{1}{c}\dfrac{\partial E'_z}{\partial t} + \dfrac{\partial H'_x}{\partial y} = -ij_z^e \\[2mm] \dfrac{1}{c}\dfrac{\partial H'_x}{\partial t} + \dfrac{\partial E'_z}{\partial y} = ij_x^m \end{cases}, \quad (5.12') \qquad \begin{cases} \dfrac{1}{c}\dfrac{\partial E'_x}{\partial t} + \dfrac{\partial H'_z}{\partial y} = -ij_x^e \\[2mm] \dfrac{1}{c}\dfrac{\partial H'_z}{\partial t} + \dfrac{\partial E'_x}{\partial y} = ij_z^m \\[2mm] \dfrac{1}{c}\dfrac{\partial E'_z}{\partial t} - \dfrac{\partial H'_x}{\partial y} = -ij_z^e \\[2mm] \dfrac{1}{c}\dfrac{\partial H'_x}{\partial t} - \dfrac{\partial E'_z}{\partial y} = ij_x^m \end{cases}, \quad (5.12'')$$

where in our case the $z$-components of the currents are equal to zero. (Note that the electromagnetic fields $(E'_x, E'_z, H'_x, H'_z)$, which define the $\psi$- function, are different than fields $(E_x, E_z, H_x, H_z)$, which define the $\Psi$ - function).

Thus, the equations (5.12') and (5.12'') are *Maxwell equations with imaginary tangential alternative currents* and simultaneously they are the Dirac-like equation with mass.

We can schematically represent the fields' motion of particles, described by these equations, in the following way (fig. 5):

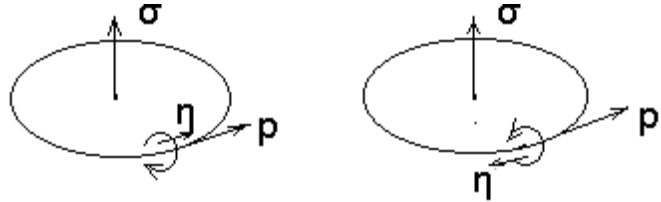

Fig. 5

According to figs. 2 and 3 the semi-photons (fig. 5) have the contrary p-helicities. In the first case the helicity vector and the Poynting vector have the same directions; in the second case they are contrary. Therefore in the non-linear theory we can define the *inner or p-helicity* as the projection of the poloidal rotation momentum on the momentum of the ring field motion.

It is not difficult to show [11] that actually the helicity is described in CWED by matrix $\hat{\alpha}_s$.

In general case for the massive particles, multiplying the Dirac equation (1.1) and (1.2) on $i\hat{\alpha}_s\hat{\beta}$ and taking in account that $i\hat{\alpha}_s\hat{\beta}\ \vec{\alpha} = \hat{\bar{\sigma}}$, where $\hat{\bar{\sigma}}' = \begin{pmatrix} \hat{\bar{\sigma}} & 0 \\ 0 & \hat{\bar{\sigma}} \end{pmatrix}$ are the spin matrix, and $\hat{\beta}\hat{\alpha}_s = -\hat{\alpha}_s\hat{\beta},\ \hat{\beta}^2 = 1$, we obtain:

$$\left(i\hat{\beta}\hat{\alpha}_5\hat{\varepsilon} + c\hat{\vec{\sigma}}'\,\hat{\vec{\rho}} - imc^2\hat{\alpha}_5\right)\psi = 0\,, \qquad (5.13')$$

$$\left(i\hat{\beta}\hat{\alpha}_5\hat{\varepsilon} - c\hat{\vec{\sigma}}'\,\hat{\vec{\rho}} + imc^2\hat{\alpha}_5\right)\psi = 0\,, \qquad (5.13'')$$

Then we can emerge for the helicity matrix the following expressions:

$$\hat{\alpha}_5 = \frac{c\hat{\vec{\sigma}}'\,\vec{p}}{i\left(\hat{\beta}\hat{\varepsilon} + mc^2\right)}, \qquad (5.14')$$

and

$$\hat{\alpha}_5 = \frac{-c\hat{\vec{\sigma}}'\,\vec{p}}{i\left(\hat{\beta}\hat{\varepsilon} - mc^2\right)}, \qquad (5.14'')$$

that in the case $m = 0$ connects the $\hat{\alpha}_5$ matrix with helicity.

From above follows that according to our theory inside the particle the operator $\hat{\alpha}_5$ describes the poloidal rotation of the fields (Fig. 5). Remembering that according to the electromagnetic interpretation [1] the value $\psi^+\hat{\alpha}_5\psi$ is the pseudoscalar of electromagnetic theory $\psi^+\hat{\alpha}_5\psi = \vec{E}\cdot\vec{H}$, we can affirm, that in the CWED the p-helicity is the Lorentz-invariant value for the massive particles, and actually it is the origin of the parity non-conservation of the massive particles.

Let's show now, that the electromagnetic fields of a matrix (4.9) which satisfy the Dirac-like equation, are not, actually fields of an initial photon.

As is known [12], the fields of a photon are vectors and will be transformed according to elements of group (O3). The spinor fields of the Dirac equation will be transformed as elements of group (SU2). As it is shown by L.H. Ryder [12], two spinor transformations correspond to one transformation of a vector. For this reason the spinors are also named "semi-vectors" or " tensors of half rank ".

Division and twirling of the photon-like waves corresponds to transition from usual linear Maxwell equation to the Maxwell equation for the curvilinear wave with an imaginary tangential current (i.e. to the Dirac-like equation). Obviously, the transformation properties of electromagnetic fields at this transition change. As wave functions of the Dirac equation (i.e. spinors) submit to transformations of group (SU2), the semi-photon fields will submit to the same transformations.

A question arises about what semi-photon fields represent and what for do they differ from the photon fields (i.e. an electromagnetic wave). In the following paragraph we will try to answer this question.

## 6.0. Specification of neutrino-like particle structure

*Let us suppose that a neutrino-like particle of above Dirac-like equation is the twirled half-period of a photon-like particle.*

In this case neutrino as twirled helicoid represents the Moebius's strip: its field vector (electric or magnetic) at the end of one coil passes to a state with opposite direction to the initial vector, and only by two coils, the vector comes to the starting position (see fig. 6)

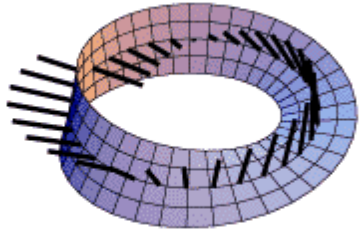

Fig. 6

(see also fig. of the Moebius's strip from [13], where the animation shows a series of gears arranged along a Möbius strip as the electric and magnetic field vectors motion)

Strict verification of the above conclusion about neutrino-like particle structure follows from the analysis of transformation properties of the twirled semi-photon wave function. Taking in account that we have the same mathematical equations, we can affirm that these transformation features coincide with the same features of the spinor [8,12].

The spinor transformation has the form:

$$\psi' = U\psi , \qquad (6.1)$$

where the operator of transformation is entered as follows:

$$U(\vec{n}\theta) = \cos\frac{1}{2}\theta - i\vec{n}\cdot\vec{\sigma}'\sin\frac{1}{2}\theta , \qquad (6.2)$$

where $\vec{n}$ is the unit vector of an axis, $\theta$ is a rotation angle around this axis and $\vec{\sigma}' = \left(\sigma_x', \sigma_y', \sigma_z'\right)$ is the spin vector.

The rotation matrix (6.2) possesses a remarkable property. If the rotation occurs on the angle $\theta = 2\pi$ around any axis (therefore occurs the returning to the initial system of reference) we find, that $U = -1$, instead of $U = 1$ as it was possible to expect. Differently, the state vector of system with spin half, in usual three-dimensional space has ambiguity and passes to itself only after turn to the angle $4\pi$ (which accords here to the one wave length of electromagnetic wave).

From above it follows that semi-photon can appear only in CWED, and in classical linear electrodynamics it do not exist.

Let us show now that such particle has mass, but it is chargeless.

## 7.0. Particle charge and mass calculations in CWED

The particle mass is defined by integral from energy density, which is proportional to the second power of the field strength. In this case the integral is always distinct from zero if the field is distinct from zero.

At the same time the charge is defined by integral on some volume from a current density, which is proportional to the first power of field strength. Obviously, it is possible to have a case when the subintegral expression is not equal to zero, but the integral itself is equal to zero. It is easy to check that we will receive such a result in case when subintegral function changes according to the harmonious law.

## 7.1. Charge calculation

It is not difficult to calculate the charge density of the twirled semi-photon-like particle:

$$\rho_p = \frac{j_\tau}{c} = \frac{1}{4\pi}\frac{\omega_p}{c}E = \frac{1}{4\pi}\frac{1}{r_p}E\,, \tag{7.1}$$

The full charge of the twirled semi-photon-like can be defined by integrating

$$q = \int\limits_{\Delta\tau_t} \rho_p d\tau\,, \tag{7.2}$$

where $\Delta\tau_t$ is the volume of fields.

Using the model (fig. 5) and taking $\vec{E} = \vec{E}(l)$, where $l$ is the length of the way, we obtain:

$$q = \int\limits_{S_t}\int\limits_0^{\lambda_p}\frac{1}{4\pi}\frac{\omega_p}{c}E_o \cos k_p l\ dl\ ds = \frac{1}{4\pi}\frac{\omega_p}{c}E_o S_c \int\limits_0^{\lambda_p}\cos\ k_p l\ dl = 0\,, \tag{7.3}$$

(here $E_o$ is the amplitude of the twirled photon wave field, $S_c$ - the area of torus cross-section, $ds$ is the element of the cross-section surface, $dl$ - the element of the length, $k_p = \frac{\omega_p}{c}$ - the wave-vector).

It is easy to understand these results: because the ring current is alternate, the full charge is equal to zero.

## 7.2. Mass calculation

To calculate the mass we must calculate first the energy density of the electromagnetic field:

$$\rho_\varepsilon = \frac{1}{8\pi}\left(\vec{E}^2 + \vec{H}^2\right), \tag{7.4}$$

In linear approximation we have $\left|\vec{E}\right| = \left|\vec{H}\right|$ in Gauss's system. Then (7.4) can be written so:

$$\rho_\varepsilon = \frac{1}{4\pi}E^2\,, \tag{7.5}$$

Using (7.5) and a well-known relativistic relationship between a mass and energy densities:

$$\rho_m = \frac{1}{c^2}\rho_\varepsilon\,, \tag{7.6}$$

we obtain:

$$\rho_m = \frac{1}{4\pi\ c^2}E^2 = \frac{1}{4\pi\ c^2}E_o \cos^2 k_s l\,, \tag{7.7}$$

Using (7.7), we can write for the semi-photon mass:

$$m_s = \int\limits_{S_t}\int\limits_l \rho_m ds\ dl = \frac{S_c E_o^2}{\pi\ c^2}\int\limits_0^{\lambda_s}\cos^2 k_s l\ kl\,, \tag{7.8}$$

From (7.8) we obtain:

$$m_s = \frac{E_o S_c}{4\omega_s c} = \frac{\pi E_o^2 r_s^2}{4\omega_s c}, \qquad (7.9)$$

Thus actually in the framework of CWED there are the cases when the particle mass do not equal to zero, but the particle charge is equal to zero.

Let us list the properties of the CWED particle, described above.

1. It is a fermion, it has mass but it doesn't have any charge.

2. There are particles and antiparticles, which are distinguished only by p-helicity .

3. The CWED neutrino has all necessary invariant properties according to the theory of the weak interaction.

*Thus the particle described by present theory can actually be named neutrino-like particle.*

Note also that in CWED two neutrino-like particles with left and right poloidal helicity should correspond to one twirled circular polarized photon-like particle. This corresponds to the theory of Luis de Broglie about the neutrino nature of light [14] if we mean the twirled (nonlinear, curvilinear) photon, not the linear .

## Conclusion

As we could ensure, within the framework of the present nonlinear theory - CWED there is a neutrino-like particle possessing the internal poloidal (torus) rotation of own fields. The reason of this rotation is the circular-polarization of semi-photon, forming this neutrino-like particle.

It was shown that the internal motion of this semi-photon on a circular trajectory is described by the Dirac-like equation of a lepton with mass equal to zero, i.e. by the Weil equation. In other words, in framework of CWED the Weyl equation is the equation of internal motion of a neutrino-like particle field. As follows from solution of the Weyl equation and as we have shown above, the internal poloidal rotation (p-helisity) allows the massive neutrino to have properties of the massless neutrino. On the other hand as a "stopped" twirled electromagnetic wave, the massive neutrino-like particle is described by the Dirac-like equation with mass term.

If we identify the CWED neutrino with the neutrino of the Standard Model, we eliminate the difficulties of the theory of Standard Model with minimum alteration of the theory. Actually, all that is necessary for overcoming the difficulties is to recognize that the neutrino has an internal motion, described by the Weyl equation.